\documentclass{tlp}

\usepackage{amsmath} 
\usepackage{times}
\usepackage{url}
\usepackage{graphicx}
\usepackage{epstopdf}
\usepackage{enumerate}
\usepackage{paralist}
\usepackage{ifpdf}
\usepackage{appendix}
\usepackage{multirow}
 \usepackage{booktabs}

\newcommand{\naf}{\ensuremath{\mathrm{not}}}

\newcommand{\dlv}{\texttt{DLV}\xspace}

\newcommand{\preceqv}{\preceq^\mathsf{v}}
\newcommand{\camouv}{\tau^\mathsf{v}}
\newcommand{\preceqp}{\preceq^\mathsf{p}}
\newcommand{\camoup}{\tau^\mathsf{p}}
\newcommand{\camouc}{\tau^\mathsf{c}}
\newcommand{\camouo}{\tau^\mathsf{o}}

\newcommand{\NP}{\mbox{\rm NP}}

\newcommand{\la}{\leftarrow}

\newcommand{\wrt}[0]{with respect to}
\newcommand{\iec}[0]{i.e.,\ }

\newcommand{\egc}[0]{e.g.,\ }

\newcommand{\nop}[1]{}
\newcommand{\kato}[0]{\texttt{Kato}\xspace}

\newcommand{\head}[1]{\mathit{H}(#1)}

\newcommand{\posbody}[1]{\mathit{B^{+}}(#1)}
\newcommand{\negbody}[1]{\mathit{B^{-}}(#1)}
\renewcommand{\max}[1]{\mathrm{max}\,{#1}}
\newcommand{\subst}{\vartheta}

\newcommand{\rules}{\mathcal{R}}
\newcommand{\identity}{\tau^{\mathsf{id}}}

\newcommand{\rulesim}[2]{\sigma({#1},{#2})}
\newcommand{\progsim}[3]{\mathcal{S}_{#3}({#1},{#2})}
\newcommand{\conf}[4]{\mathcal{C}_{{#1},{#2}}({#3},{#4})}

\newtheorem{definition}{Definition}

\usepackage{xspace}

\begin{document}
\label{firstpage}

\submitted{7 February 2010}
\revised{10 April 2010}
\accepted{22 April 2010}

\title[The System Kato: Detecting Cases of Plagiarism for Answer-Set Programs]{The System Kato: Detecting Cases of Plagiarism for Answer-Set Programs}

\author[J. Oetsch, J. P{\"u}hrer, M. Schwengerer,  and H. Tompits]
{JOHANNES OETSCH, J{\"O}RG P{\"U}HRER,\and MARTIN SCHWENGERER, AND
HANS TOMPITS%
\thanks{This work was partially supported by the Austrian Science Fund~(FWF) under grant P21698.}\\
Technische Universit\"at Wien,\\
Institut f\"ur Informationssysteme 184/3,\\
Favoritenstra\ss{}e\ 9-11,
A-1040 Vienna, Austria \\
\email{\{oetsch,puehrer,schwengerer,tompits\}@kr.tuwien.ac.at}
}

\maketitle
                                              
\begin{abstract}
Plagiarism detection is a growing need among educational institutions and solutions for different purposes exist.
An important field in this direction is detecting cases of \emph{source-code plagiarism}. 
In this paper, we present the tool \kato\ for supporting the detection of this kind of plagiarism in the area of answer-set programming (ASP).
Currently, the tool is implemented for \dlv\ programs but it is designed to handle other logic-programming dialects as well. We 
review the basic features of \kato, 
introduce its theoretical underpinnings,  and discuss an application  of \kato\ for plagiarism detection 
in the context of   courses on logic programming at the Vienna University of Technology.
\end{abstract}
\begin{keywords}
answer-set programming, program analysis, plagiarism detection
\end{keywords}

\section{Introduction}\label{sec:intro}

With the rise of the Internet and its easy access to information, plagiarism is a growing problem not only in academia but also in science and technology in general.
In software development, plagiarism involves copying (parts of) a program
without revealing the source
where it was copied from. The relevance of plagiarism detection for conventional 
program development is well acknowledged~\cite{clough00}---it is not only motivated by an
academic setting to prevent students from violating good academic standards, but
also by the urge to retain the control of program code in industrial software development 
projects.

In this paper, we deal with plagiarism detection in the context of answer-set programming (ASP), an important logic-programming formalism for declarative problem solving.
We discuss the main features and formal underpinnings of the system \kato, a tool developed for supporting the detection of source-code plagiarism for answer-set programs.\footnote{The name of the tool derives, with all due acknowledgements, from Inspector Clouseau's loyal servant and side-kick, Kato.}  
Currently, the tool is realised for logic
 programs adhering to the syntax supported by the well-known ASP solver \dlv, but it is designed to handle other syntactic dialects  as well.\footnote{%
See \url{http://www.dlvsystem.com/} for details about \dlv.}

Programming in ASP is characterised by the feature that solutions of encoded problems are determined by certain models (the ``answer sets'') of the corresponding programs. 
Moreover, the declarative nature of logic programs also marks a notable difference to 
imperative languages like C++ or Java: 
a logic program is an executable  specification rather than an  instruction on how to solve a problem;
the order of the rules and the order of the literals within the heads and bodies of the 
rules do not affect  the semantics of a program.
As well, logic programs are devoid of any control flow.
Hence, as far as plagiarism detection is concerned, someone who copies code has other means to 
disguise the deed.
Consequently, plagiarism detection tools for
imperative programming languages, 
like, e.g., \texttt{YAP3}~\cite{verco96}, \texttt{Sim}~\cite{gitchell-tran99}, \texttt{JPlag} \cite{prechelt-etal00}, \texttt{XPlag}~\cite{arwin06}, and others \cite{jones01}, are
not adequate for ASP and thus dedicated methods are needed.

The need for tools for plagiarism detection in ASP
can be motivated by the growing application in academia and industry, but our primary interest to have such a tool is to use it in connection with laboratory courses on logic programming and knowledge-based systems at our university each involving more than 100 students.
Since a manual inspection of student programs resulting from such a large body of participants is rather time consuming, we developed the tool \kato to support our grading efforts.
\kato\ implements program-based features, like string-based comparisons of  program comments, string-based comparisons of entire program sources, 
 fingerprint tests, and structure-based comparisons of programs, as well as
a context-dependent confidence measure regarding suspicious code.

A plagiarism detection system applicable in the realm of declarative programming is \texttt{Match}, implementing front-ends for Prolog and SML and following a general approach for plagiarism detection~(\citeANP{luk-sze05} \citeyearNP{luk-sze05,Lukacsy}). 
For Prolog, \texttt{Match} basically compares the  program structure given by a  program's call graph.
Although this approach can be readily adopted for ASP, the call graph of a program as discrimination criterion is somewhat too weak for the particular ASP setting we are considering.
A more detailed discussion of the {\tt Match} approach and its relation to ASP and our setting, as well as a discussion of a related idea to
compute similarities between predicate definitions due to \citeN{serebrenik}, is given in Section~\ref{sec:background}.

\section{Preliminaries on answer-set programming}\label{sec:prel}

We are concerned with \emph{disjunctive logic programs under the answer-set semantics}~\cite{gelf-lifs-91}, a widely used realisation of the ASP paradigm, consisting of 
rules of form
\begin{equation}\label{eq:rule}
a_{1} \vee \cdots \vee a_{k} \la a_{k+1}, \ldots, a_{l}, \naf\ a_{l+1}, \ldots, \naf\ a_{m}\,, 
\end{equation}
where all $a_{i}$ are literals, 
\iec atoms possibly preceded by the symbol for classical negation $\neg$,
over a function-free first-order language
and ``$\naf$'' denotes default negation.
The set of all rules of form~(\ref{eq:rule}) is denoted by $\rules$.
Given a rule $r$ of form~(\ref{eq:rule}), we define the \emph{head} of $r$ as
$\head{r}=\{a_{1}, \ldots, a_{k} \}$, the \emph{positive body} as $\posbody{r}=\{a_{k+1}, \ldots, a_{l}\}$, and the \emph{negative body} as $\negbody{r}=\{a_{l+1}, \ldots, a_{m}\}$.
We call $r$ a \emph{constraint} if $\head{r}=\emptyset$ and $\posbody{r}\cup\negbody{r}\neq\emptyset$.

\begin{figure}
\hrule
\medskip
\begin{center}
$\begin{array}{r@{~}c@{~}l}
P & =  & \left \{\ 
\begin{minipage}{10cm
}
{
\footnotesize
\begin{verbatim}
s(C,o1,X) :- combi(C), s(C,i1,X), not ab(C).
t(C,o1,X) :- combi(C), t(C,i1,Y), d(C,i2,Z), s(C,o1,1), 
             d(C,i3,heat), A = Z * 2, X = Y + A, Y <= 40, 
             not ab(C).
\end{verbatim}
}
\end{minipage} \right .\\
\\[-.2em]
Q & =  & \left\{\ 
\begin{minipage}{9cm
}
{\footnotesize
\begin{verbatim}
t(X,o1,Y) :- not ab(X),  d(X,i3,heat), d(X,i2,A), 
     s(X,o1,1), t(X,i1,Z),
     40 >= Z,  Y = Z + B, 
     B = 2 * A,  c(X).

s(X,o1,Y) :- not ab(X),
     s(X,i1,Y), c(X).
c(X) :- combi(X).
\end{verbatim}
}
\end{minipage} \right .
\end{array}$
\medskip
\hrule
\caption{Example of a program $P$ and a  disguised copy $Q$.}\label{fig:example}
\end{center}
\end{figure}

The \emph{answer sets} of a program are defined by a fixed-point construction involving the \emph{Gelfond-Lifschitz reduct} of a program \cite{gelf-lifs-91}; we omit details because we are interested in syntactic issues only.
Important to note, however, is that the order of literals in a rule, as well as the order of rules in a program, is not relevant for the semantics. 

Besides the basic syntax of rules as described above, the language of the system \dlv, which \kato\ is able to process,   
contains also \emph{built-in  predicates} for comparisons and basic integer arithmetics:
 {\tt =}, {\tt +}, {\tt *}, and 
{\tt !=}. For  {\tt !=}, the alternative syntactic form {\tt <>} is supported. 
Further built-in predicates for comparisons that are supported by $\dlv$ are
{\tt <=}, {\tt <}, {\tt >=}, and {\tt >}.
Moreover, \dlv supports \emph{aggregate functions} to express certain properties of sets, like $\tt \#sum$  or $\tt \#count$, and \emph{weak constraints} 
that allow to represent optimisation problems.

The two programs in Figure~\ref{fig:example} illustrate, on the one hand, the syntax of \dlv\ and, on the other hand, an attempt to disguise a copied program in ASP: $Q$ is a modified version of $P$ resulting from the latter by a combination of permuting and renaming expressions in
$P$, rewriting  arithmetic expressions to equivalent ones,  and  adding the  auxiliary predicate {\tt c}.
We will use $P$ and $Q$ as running example in the remainder of this paper.

\section{Background and related work on plagiarism detection}\label{sec:background}

\subsection{Text and program plagiarism}
A multitude of different approaches towards plagiarism detection for documents
exists in the literature~\cite{maurer2006plagiarism}. 
The  two main application areas of text-based plagiarism detection are
\emph{literature} (or \emph{plain text}) \emph{plagiarism} and \emph{program} (or \emph{source-code}) \emph{plagiarism}.
Interestingly, program plagiarism detection has a  longer tradition than literature plagiarism,
dating back to the 1970s~\cite{ottenstein76},
whereas extensive research on literature plagiarism detection started in the late 1990s~\cite{austinbrown99,Farringdon96analysingfor}.
The reason for this is because, on the one hand,  computer programs are well structured
and therefore easier to analyse and to compare than natural language
and, on the other hand, the  interest in literature plagiarism
is related to the boost of plagiarism that came with
the increased availability of information on the Internet.

In both cases, the key task is determining whether a given document 
is similar to an original work.
It is not sufficient for plagiarism detection systems to check for  exact copies only---they need to account
for different camouflage strategies. 
To this end, they employ diverse techniques that
provide a similarity measure with respect to some metric,
some of which are tailored towards the language of the considered document 
whereas others are language independent.
Methods that require knowledge of the language include the
comparison of writing style or frequency of spelling errors in literature plagiarism or checking similarities in the control flow in program plagiarism (when imperative languages are considered).

Among language independent methods are basic string similarity checks 
such as \emph{greedy string tiling}~\cite{greedy93} or \emph{longest common subsequence} (\emph{LCS}) \emph{tests}~\cite{bergroth00} which
are implemented in many plagiarism detectors and which can be used for plain text as well as for  programming languages.

In program plagiarism detection, two general kinds of similarity tests for programs are distinguished:
\emph{fingerprint tests} and \emph{structure tests}~\cite{whale90}.
Fingerprint-based (or \emph{attribute-based}) methods search for copies by comparing characteristic attributes
such as the numbers of unique and total operators,
which provide the basis for the program similarity metrics by \citeN{Halstead1977}.
Structure tests detect similarities based on the actual content and layout of the compared programs.
That most modern tools are mainly based on structure-based techniques~\cite{VercoWise96,Mozgovoy08} suggests
that they are  more accurate than fingerprint tests in certain applications.

A common pattern for the realisation of structure tests is applying string-similarity tests after
a preprocessing step, where the program representation is harmonised \wrt\ certain criteria.
An important preprocessing technique in this respect is \emph{tokenisation},
where the source code is translated into a token string such that certain code strings are replaced by generic tokens.
The resulting token strings are then used for further comparisons by searching for common substrings. However, 
the structure of a \dlv\ program is rather homogeneous---there are not many built-in predicates---which makes this technique rather unsuitable for detecting copies.

In general, plagiarism detection tools can be classified into systems that
operate within a given \emph{corpus}, \iec a collection of documents to check for plagiarism,
and systems that check documents also with external sources such as the World Wide Web
or textbooks \cite{lancaster-classifications}.
Plagiarism detection for programming assignments is typically done intra-corpal,
as the problems to solve are usually very specific such that it is hard for 
students to find adequate external solutions, thus they copy work from their peers.

\subsection{Plagiarism detection  for logic programs}
We next discuss structure-based approaches suitable for declarative languages and relate them to our ASP setting.
To start with, \citeANP{luk-sze05}~\citeNN{luk-sze05,Lukacsy} introduced a general framework for plagiarism detection that is applicable for
both procedural and declarative languages. 
There, the basic idea is to translate  source programs into a suitable formal representation and to apply similarity measures defined on these abstract representations. 
The system {\tt Match} instantiates this framework for, among others, Prolog  as source language.
A logic program $P$  is translated into its \emph{call graph}, also known as \emph{predicate-dependency graph}, which is 
a directed graph  where the nodes are the predicate symbols in $P$, and there is a directed edge between any  two
nodes $a$ and $b$ whenever $P$ contains a rule with $a$ in its head and $b$ in its body.
Then, the similarity between two programs is assessed by computing a graph similarity measure based on graph isomorphism
between the corresponding graph representations. 
The framework also allows to rise the level of abstraction by certain reduction steps on the graph representation. 

To use a dependency-graph representation for programs along with graph-theoretic similarity measures is certainly
an elegant way to counter many common plagiarism covering tricks. 
Changing names of identifiers and variables, changing the arity of  predicates by introducing dummy parameters, or
reordering rules and predicates in rule bodies 
are countered simply by the abstraction due to the graph translation. Other tricks like adding useless rules  or putting
some auxiliary or intermediate definitions into a program are reflected by respective structural properties of the graph representations.

In principle, the framework implemented by {\tt Match} can be adapted for answer-set programs as well. 
However, we follow an alternative approach 
due to certain particularities of ASP and the 
setting we are interested in. 
More specifically,  ASP encodings tend to be quite concise in general, and especially the programs we had to check in the context of our courses on logic
programming (the primary application area of our tool) consist of few rules only. Moreover, the language and the 
programs are rather rigorously specified. 
Hence, the structure of the dependency graph is almost identical for most pairs of programs which disqualifies the form of program abstraction used in {\tt Match} for our purposes.   
Instead, we address the similarity between two programs at the \emph{rule level}, where rules in one program are matched with
 similar rules in the other program, and a rather fine-grained control over the abstraction level of this
rule-matching procedure is introduced.
 
A further approach that can be adapted for plagiarism detection for logic programs is one by \citeN{serebrenik} who investigated methods to measure the similarity between two predicate definitions.
Interestingly, the initial motivation of this work was to detect and eliminate duplicated code that possibly resulted from refactoring steps.
In that approach, two predicate definitions are considered to be similar if they have the same recursive structure modulo renamings and 
permutations of argument positions. 
The actual  similarity  of two predicate definitions is computed, put simply,  as the sum of corresponding clause similarities that are calculated  
recursively  for the involved body predicates and reflect the amount
of common structure that can be preserved when using suitable  generalisations.
To ease the comparison of predicates, the authors introduce a fingerprinting technique, where a fingerprint is an abstraction of the recursive 
structure of a predicate. 
Thus,  this  notion of predicate similarity accounts for consistent 
renamings and permutations and
seems to be promising  for plagiarism detection for declarative logic-based languages as well.

A possible shortcoming of that approach when used for plagiarism detection is, however, that the notion of similarity between two predicates is rather semantic than syntactic.
That is to say, the form of two definitions can be quite different although their similarity is quite high. 
This is adequate for detecting duplicated code but we aim for a similarity notion that is closer tied to
the  syntactic form of the two programs that are compared.

\section{The System \kato: Architecture and basic features}\label{sec:impl}

\begin{figure}[t]
\begin{center}
\includegraphics[width=9.5cm]{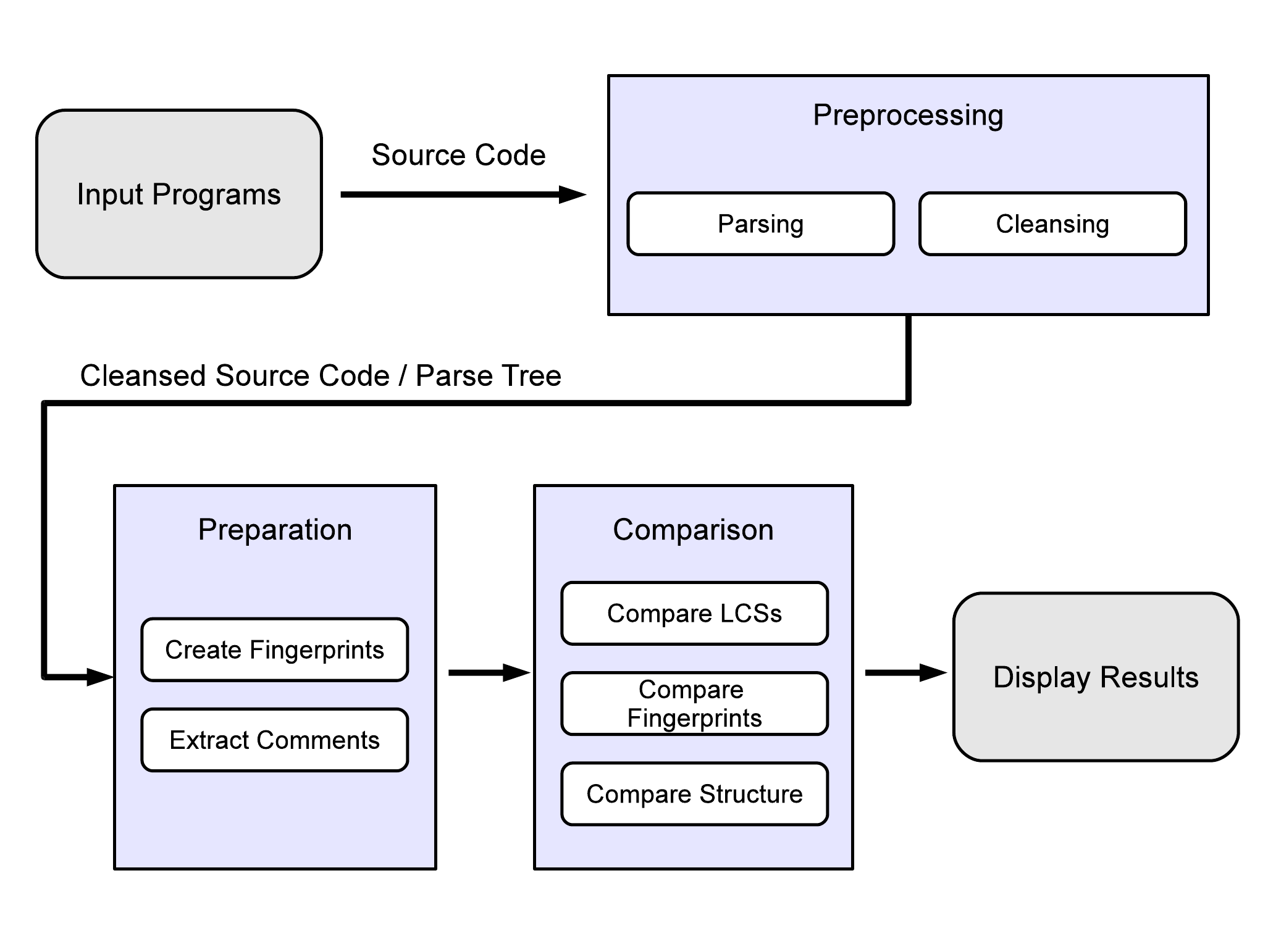}
\end{center}
\caption{Overview of how programs are  compared in \kato.}
\label{fig:kato}
\end{figure}

\kato\ was developed to find pairs of programs which are suspicious \wrt\ plagiarism. The considered programs stem
from student assignments from a course on logic programming at the Vienna University of Technology. \kato\ thus  can perform pairwise similarity tests on  rather large collections of 
 programs. 
 In what follows, we provide basic information
concerning the implemented features of \kato\ and how they are realised. 

The system was entirely developed in Java (Version 6.0) and
 provides a {graphical user interface} for controlling
test runs (a test run is a  series of pairwise program comparisons on collections of programs or single program pairs).
The user can modify the applied tests and configure how the system displays results. 
The results of a test run can be saved in a file or exported into a MySQL database.
Test results  are displayed in a table where each line gives the results of  the pairwise program comparisons. For  results of single tests, \kato\ provides 
detailed views regarding the computed similarities.
Additional information about the tool 
can  be found at
\begin{center}
\url{http://www.kr.tuwien.ac.at/research/systems/kato}.
\end{center}

Following a hybrid approach, \kato\ performs four kinds
of comparison tests, 
realising different layers of granularity:
(i)~a string-based comparison of the program comments, 
(ii)~a string-based comparison of entire program sources, 
(iii)~a fingerprint test, and 
(iv)~a structure-based comparison of  programs. 

\begin{figure}[t]
\begin{center}
\includegraphics[width=\textwidth]{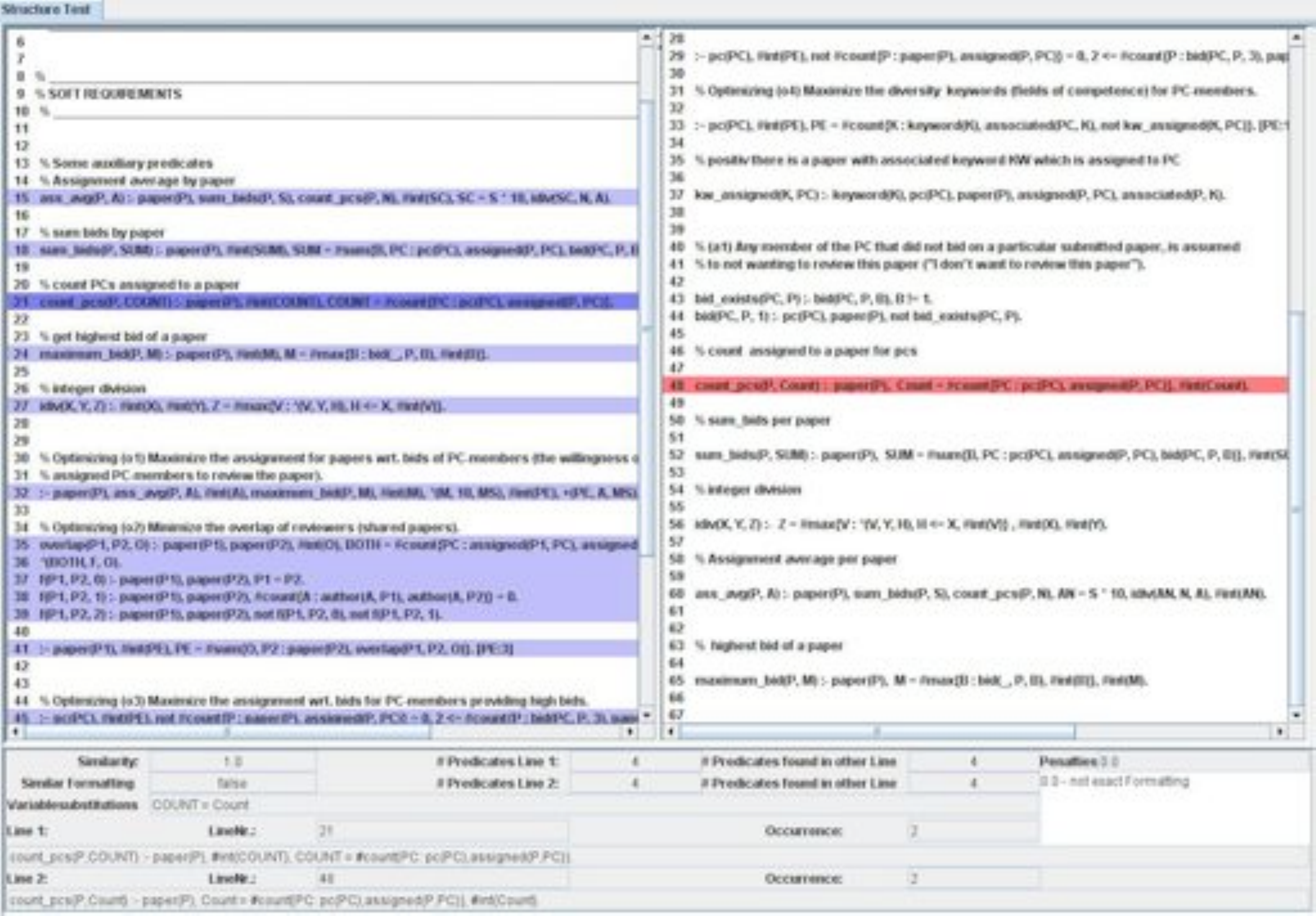}
\end{center}
\caption{A screen-shot showing how \kato\ displays the results of  a structure test. }
\label{fig:screenshot}
\end{figure}

For the string-based tests, comparisons based on a {longest common subsequence} (LCS) metric are applied \cite{bergroth00}.
Long common substrings are usually an indicator for a copy, however a major drawback of this method is its vulnerability \wrt\  non-matching  symbols in a sequence which break the substrings apart~\cite{LBPS08}.
On the other hand, as the LCS of two strings is the longest sequence of symbols appearing in both strings with the same succession, the LCS metric tolerates injected non-matching objects and is more robust in this respect.
To illustrate the notion of LCS, consider the two strings
``\kato\ is a great tool for plagiarism detection in ASP'' and
``the system \kato\ is currently applied in a course on logic programming to reveal cases of plagiarism''.
The LCS of these strings is ``\kato is a plagiarism''.
The \emph{LCS similarity} of two strings $s$ and $t$ is the 
 length of the LCS of $s$ and $t$ normalised by the length of $s$.

We note that tests (i) and (ii) are language independent whilst (iii) and (iv) need to be adapted for
different languages.
All of these tests, outlined in more detail below, compare files pairwise and return a similarity value between 0 (no similarities) and 1 (perfect match). 

Figure~\ref{fig:kato} shows the basic working steps needed to compare programs in \kato.
First, the source code  of a collection of  programs is
cleansed, \iec white spaces are removed, and parsed  in a  preprocessing step. 
Then, the cleansed source code, as well as the abstract program representations resulting from the parsing step, are
 subject to a preparation step where fingerprints are created and comments are extracted. 
In a comparison phase, LCS comment tests, LCS program tests, fingerprint tests, and structure tests are applied. 
Finally,
the results of the comparisons are presented to the user. 
More specifically, the results are displayed in 
tabular form with features like sorting and filtering.
For the structure tests, the tool shows program pairs and 
highlights similar rules; see Figure~\ref{fig:screenshot} to get a rough impression. 

In what follows, we describe the comparison tests (i)--(iv) in more detail.

\paragraph{LCS comment test.}
When programs are copied, it is surprisingly quite frequent that program comments are copied as well and usually little effort is spent to mask such copied comments.
Consequently, the LCS comment test aims at revealing similar parts in program comments.
More specifically, this test reveals similarities between two programs by comparing the (concatenated) comments
in the two programs via the LCS similarity measure.

\paragraph{LCS program test.}
This test is very similar to the LCS comment test but considers entire programs.
Thus, this test computes the LCS similarity of the  cleansed source code of two programs when interpreted as strings.
It turned out that this test is an efficient method to detect
cases of plagiarism where not much time has been spent to camouflage the deed.  
It is  especially well-suited to detect perfectly matching sequences in programs which is usually a
clear hint for plagiarism.

\paragraph{Fingerprint test.}
Recall that a fingerprint of a program is a collection of relevant program attributes, \iec statistical data like 
hash codes, the number of rules, the number of predicates, the number of constants, program size, and so
on. After fingerprints of all programs are generated in a preparation step, the fingerprints are compared
pairwise. 
This test is quantified by normalising the number of matching attributes by the total number of considered attributes for each pair
of programs.
This gives a simple yet convenient way to collect further evidence for plagiarism.

\paragraph{Structure test.}
This test does not operate on the cleansed program sources but on the program representation that was built in
a preceding parsing step.
We here rely on a set representation of programs and of rule heads and bodies as well as on a structured representation of literals and term lists.
While the LCS comment test, the LCS program test, and the fingerprint test are, more or less, standard techniques to detect plagiarism in general, the structure test is developed specifically for our ASP setting.

The central notion of \emph{program similarity}  that underlies  
the structure test,  formally defined in the next section,
is designed to thwart disguising strategies like permuting rules or literals
within rules. 
However, a more advanced plagiarist will apply additional camouflage techniques, \egc  uniformly renaming variables in rules or renaming auxiliary predicates in programs. 
Therefore, our similarity test comes with different levels of abstraction, so-called \emph{camouflage techniques}, to counter such efforts.
Without going into details at this point, renaming is handled by finding and applying suitable substitution functions.

To make the similarity function sensitive to common rule patterns, we also implemented a context dependent notion 
of \emph{confidence}
regarding plagiarism:
When whole collections of programs are examined for suspicious pairs of programs,
a \emph{global occurrence table} gives
additional information on how specific two rules are. The main idea is that rare rules yield better evidence for a copy than common ones. Therefore, \kato\ collects and counts all rules
in the considered corpus of programs
and stores this information in an {occurrence table} which is then used to compute a measure of confidence regarding
plagiarism for two programs based on the frequency of common rules in the programs relative to their frequency in the
considered corpus. 

\section{The formal similarity model of the structure test}\label{sec:theory}

In this section, we present a syntactic measure of program similarity between logic programs that forms the theoretical basis of the central plagiarism-detection features of \kato.
Furthermore, we introduce a  confidence measure that aims at quantifying the  reliability  of our similarity measure for detecting actual cases of plagiarism in ASP. 

\subsection{A syntactic measure for rule and program similarity}

Our basic assumption is that a plagiarist camouflages a copy by changing the form of the copied program, \iec by \emph{program transformations at a syntactic level}.  
Examples for rather simple camouflage techniques  are changing the formatting of the source code,
permuting rules of the program, and changing the order of literals within rule heads or bodies. 
More advanced techniques include uniformly renaming variables within rules, renaming auxiliary atoms
within programs, and rewriting   arithmetic expressions to equivalent ones.  
Clearly, all these basic techniques can be combined to build more complex operations.

A key concept for detecting copied programs is thus measuring the similarity between programs. 
Formally, this boils down to the definition of a suitable similarity function that quantifies the syntactic similarity between two programs  with respect to specific combinations of camouflage techniques. 
To cover different camouflage techniques in a uniform way and to allow for the composition of different techniques, we handle them as functions that map pairs of rules to pairs of rules.  

\begin{definition}\label{def:technique}
A \emph{camouflage technique}, or \emph{technique}, is a function
with domain and co-domain $\rules \times \rules$. 
In particular, by the \emph{identity technique}, $\identity$, we understand the identity function over $\rules \times \rules$, \iec $\identity$ satisfies $\identity(p)=p$, for each $p \in \rules \times \rules$.
\end{definition} 

Obviously, the notion of camouflage technique is closed under functional composition, \iec
for any two techniques $\tau_{1}$ and $\tau_{2}$, the composition $\tau_{1} \circ \tau_{2}$ is
a technique as well.
From an algebraic point of view, the set of techniques and the composition operator form a monoid  with
$\identity$ as its identity element. 
Hence, the above definition allows to express more complex combinations of different techniques, as used, \egc to disguise program $Q$ 
in Figure~\ref{fig:example}, by a composition of less complex  techniques. 
  
We next introduce our central similarity measures.
Roughly speaking, we define the similarity between two single program rules $r$ and $s$ as
the number of common head and body literals of $r$ and $s$ normalised by the total number of
literals in $r$. 

\begin{definition}\label{def:rulesim}
For any two rules $r,s \in \rules$, 
the \emph{rule similarity between $r$ and $s$} is given by
\[
\rulesim{r}{s} = 
\frac{{|\head{r} \cap \head{s} |+ |\posbody{r} \cap \posbody{s}|  + |\negbody{r} \cap \negbody{s}|}}
{|\head{r}| + |\posbody{r}| + |\negbody{r}|}.
\]
\end{definition}

For example, consider the two rules
\begin{center}
\begin{tabular}{l@{}c@{}l}
$r$ & $=$  &{\tt \small p(X) :- r(X), s(X,Z), not t(c)} \ and \\
$s$ & $=$ & {\tt \small p(X) v q(Y) :- not\ t(c), s(X,Z), r(Y)} .
\end{tabular}
\end{center}
According to the above definition, $\rulesim{r}{s} = \frac{3}{4}$.

The notion of rule similarity extends to entire programs as follows:

\begin{definition}\label{def:progsim}
Given two programs $P_{1}$ and $P_{2}$ together with a camouflage technique $\tau$,
 the \emph{program similarity between $P_{1}$ and $P_{2}$ \wrt\ $\tau$} is given by
\[
\progsim{P_1}{P_2}{\tau} = \frac{\sum_{r \in P_1}  \max{\{\sigma\big(\tau(r,r')\big) \mid r' \in P_2\}}}{|P_1|} .
\]
\end{definition}

Clearly, for any program $P_1$ and $P_2$ and for any technique $\tau$, it holds that
$\progsim{P_{1}}{P_2}{\tau}\in[0,1]$.
Note that rule similarity and  program similarity are not symmetric in their arguments.
 Roughly speaking, the significance of this asymmetry is that it allows to express
to which extent $P_1$ is subsumed by $P_2$ by similar rules \wrt\ $\sigma$ and $\tau$.
This can be further interpreted as hints concerning to which extent one program resulted from another program
by adding or splitting certain rules which can help to assess who copied from whom.

Using the  identity technique to the two programs $P$ and $Q$ from
Figure~\ref{fig:example} to compute its similarity yields
$\progsim{P}{Q}{\identity} = \progsim{Q}{P}{\identity} = 0$. 
Hence, we need more advanced techniques than
$\identity$ to get a similarity function of practical use. 
Accordingly, we next introduce some basic techniques which can serve as building blocks for more complex ones.

\subsection{Basic camouflage techniques}\label{subsec:techniques}

\subsubsection{Variable renaming}

The first technique we consider is the uniform renaming of variables within rules which
is a  simple way to change the form of a rule. 
Formally, a \emph{variable renaming} for a rule $r$ is a bijection 
from the set $S$ of variables occurring in $r$ to a set $S'$ consisting of $|S|$ arbitrary variables.
For any variable renaming $\subst$  for some rule $r$, $r\subst$ denotes the result of 
replacing each occurrence of a variable $x$ in $r$ by $\subst(x)$. 
We assume that $\preceqv$ is a globally fixed well-ordering on variable renamings
that  extends to tuples of variable renamings in a lexical way.
Informally, the following technique applies variable renamings to a pair of rules that results in a maximal 
rule similarity. Since such maximal variable renamings are not unique in general, we need  $\preceqv$ to define
a proper function.

\begin{definition}\label{def:varrenaming}
The camouflage technique $\camouv$ is the function assigning to
each rule pair $(r,s)\in \rules\times\rules$ the pair
$\camouv(r, s) = (r\subst_{r}, s\subst_{s})$,
where $\subst_{r},\subst_{s}$ are the variable renamings for $r$ and $s$ such that
\[
\begin{array}{r@{~}l}
(\subst_{r},\subst_{s})=\mathrm{min}_{\preceqv}\{(\subst_{r}',\subst_{s}')\mid & \rulesim{r\subst_{r}''}{s\subst_{s}''} \leq \rulesim{r\subst_{r}'}{s\subst_{s}'}\mbox{, for all variable renamings}\\
&\mbox{$\subst_r''$ and $\subst_s''$ for $r$ and $s$ such that }(\subst_{r}'', \subst_{s}'')  \neq (\subst_{r}', \subst_{s}')\}.
\end{array}
\]
\end{definition}

For example, consider programs $P$ and $Q$ from Figure~\ref{fig:example}.  Let $r$ be the last rule  in  $P$ and $s$ the first 
rule in $Q$.
Rule $r$ contains the variables $C$, $X$, $Y$, $Z$, and $A$, and $s$ contains the variables $X$, $Y$, $Z$, $A$, and $B$. 
Variable renamings  that yield maximal rule similarity are
\(\subst = \{
C \mapsto X,
X \mapsto Y,
Y \mapsto Z,
Z \mapsto A,
A \mapsto B
\}\) for $r$ and the identity function for $s$. 

Applying $\subst$ to $r$ results in the rule
\[
\begin{array}{r@{~}c@{~}l}
      \texttt{\small t(X,o1,Y)}& \texttt{\small :-} & \texttt{\small combi(X), t(X,i1,Z), d(X,i2,A), s(X,o1,1),}\\ 
& & \texttt{\small d(X,i3,heat), B = A * 2, Y = Z + B, Z <= 40, }\\
&& \texttt{\small not ab(X).}
\end{array}
\]
Assuming a suitable well-ordering, we get $\camouv(r,s) = (r\subst,s)$, and thus
$\sigma(\camouv(r,s)) = \frac{7}{10}$.
Note that the similarity between $r$ and $s$ under $\identity$ is  $0$. 

\subsubsection{Predicate renaming}
Next, we address a technique to deal with renaming efforts at the predicate level. 
Given a program $P$,  a bijection $\subst$ from the set $S$ of predicate symbols occurring  in $P$  to  a set $S'$
consisting of $|S|$ arbitrary predicate symbols is a \emph{predicate renaming} for $P$ if
$p$ and $\subst(p)$ have the same arity, 
for each $p \in S$.
Let $\subst$ be a predicate renaming for a program and $E$
a program or a rule. Then, $E\subst$ denotes the result of 
replacing each occurrence of a predicate symbol $x$ in $E$ by $\subst(x)$, provided $\subst$ is defined for $x$.  

Note that predicate renamings are not applied to single rules but to entire programs.
To express predicate renamings in terms of camouflage techniques, we 
introduce a  technique that is parameterised by two programs that are used to determine
 the predicate renaming that is applied.\footnote{Formally, this means that we introduce a family of techniques, one technique for each  program pair, respectively.}
Like for variable renamings, we fix a well-ordering $\preceqp$ defined on predicate renamings
and assume that  $\preceqp$ extends to tuples of predicate renamings in a lexical way.

\begin{definition}\label{def:predrenaming}
Given two programs $P$ and $Q$, the camouflage technique $\camoup_{P,Q}$ is the function assigning to
each rule pair $(r,s)\in \rules\times\rules$ the pair
$\camoup_{P,Q}(r,s) = (r\subst_{r}, s\subst_{s})$,
where $\subst_{r}$ and $\subst_{s}$ are the predicate renamings for $P$ and $Q$ such that
\[
\begin{array}{r@{}l}
(\subst_{r},\subst_{s})=\mathrm{min}_{\preceqp}\{&(\subst_{r}',\subst_{s}')\mid \progsim{P\subst_{r}''}{Q\subst_{s}''}{\tau^{v}} \leq \progsim{P\subst_{r}'}{Q\subst_{s}'}{\tau^{v}}\mbox{, for all predicate}\\
&\mbox{renamings $\subst_r''$ and $\subst_s''$ for $P$ and $Q$ such that }(\subst_{r}'', \subst_{s}'')  \neq (\subst_{r}', \subst_{s}')\}.
\end{array}
\]
\end{definition}

Intuitively, the above definition formalises the idea of applying predicate renamings that result in maximal
program similarity when also variable renamings are considered for the rule comparisons. 

Recall again programs $P$ and $Q$ from Figure~\ref{fig:example}. 
Predicate renamings that yield maximal program similarity are
the renaming $\subst$ for $P$ which maps each predicate symbol in $P$ to itself except
for $\mathtt{combi}$, which is mapped to $ \mathtt{c}$, and the renaming $\subst'$ for $Q$ given as the identity function. 
 
Let us consider the technique $\tau =  \camoup_{P,Q} \circ \camouv$.
Then,
$\progsim{P}{Q}{\tau} = \frac{9}{10}=0.9$.
Observe
that $\progsim{Q}{P}{\tau'} = \frac{18}{30}=0.6$, for $\tau' =  \camoup_{Q,P} \circ \camouv$, which
illustrates that
$P$ is almost entirely subsumed (syntactically) by $Q$ but not vice versa due to the additional third rule in $Q$.

\subsubsection{Canonisation of built-in predicates}

The next technique is designed to counter program transformations that exploit that
some \dlv\ built-in predicates have  syntactically  different 
representations.
For each built-in predicate, we define a unique \emph{canonical  representation}.
E.g., the canonical representation of {\tt Z = X + Y} and {\tt +(X,Y,Z)} is {\tt +(X,Y,Z)} (the canonical representations for {\tt  *}, {\tt  =}, {\tt  <>}, {\tt  !=}, {\tt  <=}, {\tt  <}, {\tt  >=}, and {\tt  >} are omitted for space reasons).

\begin{definition}\label{def:canonisation}
The camouflage technique $\camouc$ is the function assigning to each rule pair $(r,s) \in \rules\times\rules$ the pair $\camouc(r, s) = (r', s')$,
where $r'$ results from $r$ by rewriting all occurrences of built-in predicates in $r$ into their respective canonical forms, and likewise for $s'$ and $s$.
\end{definition}

Concerning our running example from Figure~\ref{fig:example},
if we pool the techniques in our  arsenal of camouflage techniques considered so far, we get a program similarity of
$\progsim{P}{Q}{\tau} = \frac{19}{20}$, for $\tau = \camoup_{P,Q} \circ \camouv \circ \camouc$.

\subsubsection{Ordering the arguments of commutative predicates}
For most \dlv\ built-in predicates, viz.\ for {\tt =}, {\tt +}, {\tt *}, 
{\tt !=}, and {\tt <>}, (some of) their arguments can be commuted. 
Swapping arguments of such \emph{commutative predicates} is addressed by the  next technique.

\begin{definition}\label{def:swapping}
The camouflage technique $\camouo$ is the function assigning to each rule pair $(r,s) \in \rules\times\rules$ the pair
$\camouo(r, s) = (r', s')$,
where $r'$  is the rule that results from $r$  by replacing each commutative predicate $p$ in $r$
 with  a rewriting of $p$ where all commutable arguments of $p$ are lexically ordered according to their
string representations, and $s'$ results from $s$ in an analogous fashion. 
\end{definition}
 
For programs $P$ and $Q$ from  Figure~\ref{fig:example},
we eventually obtain $\progsim{P}{Q}{\tau} = 1$, for $\tau = \camoup_{P,Q} \circ \camouv \circ \camouc \circ \camouo$,
and  $\progsim{Q}{P}{\tau'} = \frac{2}{3}=0.\dot{6}$, for $\tau' = \camoup_{Q,P} \circ \camouv \circ \camouc \circ \camouo$.

\subsubsection{Further issues}
\paragraph{Customisation of techniques in {\em \kato}.}
Versions of all the basic camouflage techniques discussed so far are implemented in \kato. 
Note, however, that the composition operation for these techniques
is not commutative.
Thus, different compositions result in different similarity values between programs in general.
As well, the user can easily define further techniques from the basic ones, but our current implementation is restricted to compositions subject to the order
$
\camouo < \camouv < \camouc < \camoup_{P,Q} < \identity
$ which turned out to be particularly useful in practice.
We plan to extend \kato such that this limitation is removed in future versions.
Furthermore, \kato\ provides the following hierarchy of predefined techniques:
\begin{center}
$\tau_{1} = \identity$, $\tau_{2} = \camouv$, $\tau_{3} = \camouo \circ \camouv \circ \camouc$, and $\tau_{4} = \camouo \circ \camouv \circ \camouc \circ \camoup_{P,Q}$.
\end{center}
Each of these techniques realises a level
of abstraction: On the one hand, for two programs $P$ and $Q$, usually 
$\progsim{P}{Q}{\tau_{i}} \leq \progsim{P}{Q}{\tau_{j}}$ holds  if 
$i < j$. 
On the other hand,  the significance of a high similarity value relative to $\tau_{i}$ is indirect proportional to $i$.
It often makes sense to consider a sequence of comparisons, starting with $\tau_{1}$ and ending with $\tau_{4}$, and to consider
similarity and confidence relative to the level $i$---\kato\
can perform such hierarchical tests, if required.
We will discuss an empirical analysis employing these predefined techniques in Section~\ref{sec:eval}.

\paragraph{Complexity aspects.}
Concerning the inherent  complexity of applying a technique to a pair of rules, 
we note that the variable and predicate renaming techniques tend to be computationally expensive while
the other techniques can be applied in linear time \wrt\ the size of the rules.
 Without going into details,  the problem of finding variable or predicate substitutions is  related to the 
homomorphism problem for relational structures which is known to be \NP-complete (when formulated as a decision problem). However, high complexity
 of the 
involved reasoning tasks does not
turn out to be a bottle-neck of our approach since this worst-case complexity is relative to the size of rules or 
 programs and
problem encodings in ASP tend to 
be quite concise and rarely involve a larger number of rules---this is witnessed, \egc by the collection of benchmark programs
used for the recent ASP solver competition~\cite{competition09}.

\subsection{A corpus-based confidence measure}

We now introduce a method how to further evaluate the outcome of a program comparison regarding its
potential to actually reveal a  case of plagiarism.
One important aspect in this regard is the influence of the used camouflage technique $\tau$ on the program similarity.
In general,
we can observe a trade-off between the complexity of $\tau$ (in terms of operations on rules) and
the significance of a high similarity value for revealing plagiarism. 
As illustrated, the similarity between programs $P$ and $Q$ from Figure~\ref{fig:example} ranges from $0$ to $1$, depending on the used technique.
However, the technique used to obtain a similarity of 1 is rather demanding and presumes that 
a plagiarist applied more advanced methods to cover a copy.
Thus, the significance of the similarity outcome is rather low.

A common and in practice important setting  is when we have a collection, or corpus, of programs, and our goal is
to detect suspicious pairs of programs within this corpus.  Such a setting allows to factor in further information
that can increase our confidence that a high program similarity actually indicates a case of plagiarism.  
Here, the basic idea is to consider the relative frequency of rule occurrences \wrt\ the considered program collection.
If two programs have a high similarity but share only very common rules, our confidence 
will be accordingly low. 
Likewise, if two programs contain the same extremely rare rule,
our confidence for plagiarism will be rather high.
Clearly, the notion of ``sameness'' of a rule depends on the considered camouflage technique $\tau$.
Formally, we introduce a suitable equivalence relation based on $\tau$  and consider the induced equivalence class to define the relative frequency of a rule \wrt\ a program corpus.

Given a camouflage technique $\tau$   composed from techniques in $\{\camouv, \camoup_{P,Q}, \camouc, \camouo\}$, let us
define the  relation $\sim_{\tau}\ \subseteq \rules \times \rules$ as 
$
r \sim_{\tau} s \mbox{ iff }  \sigma(\tau(r,s)) =  \sigma(\tau(s,r)) =  1\,,
$
for any rule $r,s \in \rules$.
It can be shown that  $\sim_{\tau}$ is an equivalence relation.
For a set $S \subseteq \rules$ of rules, a rule $r \in S$, and a relation  $\sim_{\tau}$, let $[r]_{\tau}$ be the  equivalence class of 
$r$ in $S$ under $\sim_{\tau}$.
The \emph{relative frequency} of $r$ \wrt\ $S$ and $\tau$ is given by 
$$f_{S,\tau}(r)=\frac{|[r]_{\tau}|}{|S|}.$$
The notion of confidence regarding two programs is then defined as
follows:

\begin{definition}\label{def:confidence}
Let $\mathcal{P}$ be a collection of programs and $\tau$ a camouflage technique   composed from techniques in $\{\camouv, \camoup_{P,Q}, \camouc, \camouo\}$.
Furthermore, let $S$ be the set of all rules occurring in $\mathcal{P}$.
Then, for programs $P, Q \in \mathcal{P}$,
the \emph{confidence regarding $P$ and $Q$ relative to $\mathcal{P}$ and $\tau$} is given by
\[
\conf{\mathcal{P}}{\tau}{P}{Q} = \max{\{1 - f_{S,\tau}(r) \mid r \in P\mbox{ such that there is some $s \in Q$ with } r \sim_{\tau} s} \}\,.
\]
\end{definition}

\section{Application of \kato and empirical analysis}\label{sec:eval}

As mentioned earlier, our original motivation for the development of \kato\ was to have a tool to support detecting cases of plagiarism in the context of several laboratory courses.

The particular setting of the courses was as follows:
Students were assigned with exercises and had to solve them 
 by means of ASP. The size of the program parts from the student solutions ranged from a few rules to dozens of rules.
Besides evaluating the correctness of the solutions, human supervisors had
to check for plagiarism. 
Since the number of students increased from year to year, with typically over 100 students attending the courses, 
the task to manually inspect all pairs of programs for plagiarism became too time consuming and laborious.
We thus decided in early 2009 to develop \kato, which became 
operational in the summer term of that year.
 \kato\ was used to pre-select pairs of programs if their program similarity was beyond a fixed threshold, and the selected programs were then further manually inspected by the supervisors. 
\kato had to perform tens of thousands of program comparisons 
which was feasible within 
hours.
Most importantly, the application of the system yielded dramatical savings in labour time.

To provide a more detailed assessment of \kato, 
in what follows we give an evaluation of \kato in terms of the well-known notions of  \emph{precision} and \emph{recall}.
In our setting,
precision is the ratio between the number of program pairs that are correctly classified by \kato as copies 
and the total number of 
pairs that are classified by \kato as copies (\iec the number of true positives divided by the sum of true positives and false positives), and
recall is the ratio between the number of program pairs that are correctly classified as copies and the total number of pairs that are 
actual copies (\iec the number of true positives divided by the sum of true positives and false negatives).
The purpose of the evaluation is to show the effects of different combinations of techniques for the structure test on precision and recall.

For the present evaluation, we used data from the 
logic programming course that took place in 2008 where we had to assess 109 programs from the students.
We did not use assignments from the courses after 2008 because these were used as reference data in the development phase of \kato and thus are 
unsuitable for a fair evaluation as the tool is trained on them.
In the considered program collection, we manually identified 26 program pairs suspected for plagiarisms.

\begin{table}
	\caption{Evaluation of techniques for the structure test.}
	\label{tab:eval}
{\footnotesize
	\begin{center}
	\begin{tabular}{ccccccc
}
	  \hline
	  	     {Technique} & 
	  	     {Time} & 
	  	    {Threshold} & 
	  	    {Classified} & 
	  	    {Actual} & 
	  	    {Recall} & 
	  	    {Precision} \\
	    & (sec.) & & as copies & copies & & \\
	    \toprule
	  \multirow{4}{*}{$\tau_{1}$ } & \multirow{4}{*}{$12$ } & $1.00$ & $5$ & $5$ & $0.19$ & $1.00$  \\ 
	  &  & $0.95$ & $7$ & $7$ & $0.27$ & $1.00$ \\ 
	  &  & $0.90$ & $11$ & $9$ & $0.35$ & $0.82$ \\ 
	  &  & $0.85$ & $11$ & $9$ & $0.35$ & $0.82$ \\ 
	\midrule
	  \multirow{4}{*}{$\tau_{2}$ } & \multirow{4}{*}{$49$ } & $1.00$ & $5$ & $5$ & $0.19$ & $1.00$ \\ 
	  &  & $0.95$ & $7$ & $7$ & $0.27$ & $1.00$ \\ 
	  &  & $0.90$ & $11$ & $9$ & $0.35$ & $0.82$ \\ 
	  &  & $0.85$ & $11$ & $9$ & $0.35$ & $0.82$ \\ 
	\midrule
	  \multirow{4}{*}{$\tau_{3}$ } & \multirow{4}{*}{$94$ } & $1.00$ & $5$ & $5$ & $0.19$ & $1.00$ \\ 
	  &  & $0.95$ & $7$ & $7$ & $0.27$ & $1.00$ \\ 
	  &  & $0.90$ & $15$ & $11$ & $0.42$ & $0.73$ \\ 
	  &  & $0.85$ & $16$ & $11$ & $0.42$ & $0.69$ \\ 
	\midrule
	  \multirow{4}{*}{$\tau_{4}$ } & \multirow{4}{*}{$169$ } & $1.00$ & $53$ & $13$ & $0.50$ & $0.25$ \\ 
	  &  & $0.95$ & $113$ & $21$ & $0.81$ & $0.19$ \\ 
	  &  & $0.90$ & $174$ & $25$ & $0.96$ & $0.14$ \\ 
	  &  & $0.85$ & $266$ & $25$ & $0.96$ & $0.09$  \\ 
	  \bottomrule
	\end{tabular}
	\end{center}
}
\vspace{-1\baselineskip}
\end{table}

Table~\ref{tab:eval} shows the results of our evaluation of different techniques for the structure test, where
we considered the predefined techniques $\tau_{1}$--$\tau_{4}$ from above. Recall that $\tau_{1}$ is the simplest technique while $\tau_{4}$ is the most
complex one.
The table is organised as follows.
In each of the four main rows, we give the precision and recall results for one technique $\tau_{1}$--$\tau_{4}$ as well as the time to run the test.
For each technique, we considered different values, ranging from $0.85$ to $1$,  for the similarity threshold used to
classify whether a pair of programs counts as a case of plagiarism.
For each threshold value, we give the number of program pairs that are classified as copies by $\kato$, \iec that showed a program
similarity above the threshold value, as well as the number of actual copies, \iec the number of program pairs from the ones classified
 as copies by \kato that are indeed cases of plagiarism according to a human supervisor.
The last two columns depict the actual precision and recall values.

The evaluation clearly shows the trade-off between precision and recall relative to the abstraction level imposed by the considered
camouflage technique. 
For the simplest technique~$\tau_{1}$, precision is maximal and recall is minimal.
When going from the simpler techniques to
the more complex ones, precision decreases while recall increases.
This observation nicely illustrates how precision and recall can be controlled by \kato's system of composable camouflage techniques.
Furthermore, the results indicate also that each of our considered camouflage techniques is used by students, although some of them occur usually in combination with others. 
For example, students who renamed auxiliary predicates 
usually also rename variables. 
According to our evaluation, many students who change the names of auxiliary predicates also perform other changes, such as reordering commutative operators or renaming variables. 

Concerning  the different plagiarism detection strategies realised in $\kato$,
our experience suggests that the structure test provides the most accurate results when  
students spent some effort to camouflage the copy. 
The LCS program test, on the other hand, revealed  exact or almost exact copies, while the LCS comment test seems to be a good additional indicator for plagiarism. 
The fingerprint test  provided a fast way to 
 detect exact copies because of  identical hash values.
However, it is to mention that the exercises were rather rigorously specified which implies that it was rather common
that programs agreed on some attributes like, \egc the number of different predicate symbols.

 \section{Conclusion }\label{sec:concl}

 In this paper, 
 we presented the tool \kato for plagiarism detection  in ASP, reviewing its underlying  methodology and its
 basic features.
 In particular, we introduced the formal basis of \kato:   syntactic notions of program similarity between programs 
 along with a formal measure of confidence regarding
 plagiarism detection for programs. Both concepts are based on the notion of a camouflage technique which is
 basically a strategy to  disguise
 the form of a program and which provides
 a flexible means to deal with different strategies and combinations thereof in a uniform manner.
 Currently, \kato\ is designed for \dlv's language dialect but it can easily be  extended
 to other dialects.
 For \dlv programs, \kato supports also weak constraints and aggregates, and it can also deal with programs for the  planning front-end of  \dlv~\cite{eiter03}. 
The system was successfully applied for laboratory courses 
at our university and helped to save valuable labour time.  

 For future work, we plan  to develop means to visualise the comparison results, e.g.,  
 to spot clusters of cooperating plagiarists more easily.
 A further  aspect of \kato\ worth mentioning is a possible use for  the development of logic programs:
 If a team is working on a program, different versions can emerge. Then, assessing 
 the actual differences between two versions can be necessary. \kato\ can be adapted for
such an application scenario.
 

\end{document}